\documentclass{appolb}
\usepackage{epsfig}
\newcommand{\lesssim}{\raisebox{0.3mm}{\em $\, <$} 
\hspace{-3.3mm} \raisebox{-1.8mm}{\em $\sim \,$}}
\newcommand{\gtrsim}{\raisebox{0.3mm}{\em $\, >$}
\hspace{-3.3mm} \raisebox{-1.8mm}{\em $\sim \,$}}
\begin{document}
\pagestyle{plain}

\title{New Physics Effects in
Long Baseline Experiments
\thanks{Presented at the XXXI International School of Theoretical
Physics ``Matter To The Deepest:
Recent Developments In Physics
of Fundamental Interactions'', Ustron, Poland,
September 5--11, 2007.
}
}
\author{Osamu Yasuda
\address{Department of Physics,
Tokyo Metropolitan University \\
Minami-Osawa, Hachioji, Tokyo 192-0397, Japan}
}
\maketitle

\begin{abstract}
We discuss
the implications of new physics, which modifies
the matter effect in neutrino oscillations,
to long baseline experiments,
particularly the MINOS experiment.
An analytic formula in the presence of such a new physics interaction
is derived
for $P(\nu_\mu\rightarrow\nu_e)$,
which is exact in the limit $\Delta m^2_{21}\rightarrow0$.
\end{abstract}
\vskip 0.5cm

\section{Introduction}
It has been suggested that future long baseline neutrino experiments
such as so-called super-beams, beta-beams, neutrino factories
will have great sensitivity to the third mixing angle $\theta_{13}$,
the CP phase $\delta$ and the mass hierarchy
$\mbox{\rm sign}(\Delta m^2_{31})$
(For a review, see, e.g., \cite{ISS}).
Just like at the B factories, experiments of high precision measurements
will allow us not only to measure precisely
the parameters of the standard model,
but also to probe new physics by looking at a deviation from
the standard case.
In this talk I would like to discuss the possible effects
of new physics at long baseline experiments,
particularly the MINOS experiment~\cite{Michael:2006rx}.

\section{New physics in neutrino oscillations}
\begin{figure}[h]
\begin{center}
\epsfig{file=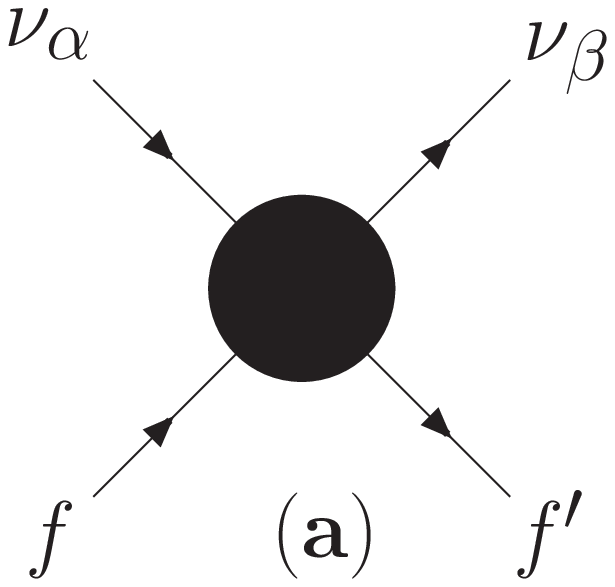,width=2.5cm}
\hglue 0.2cm
\epsfig{file=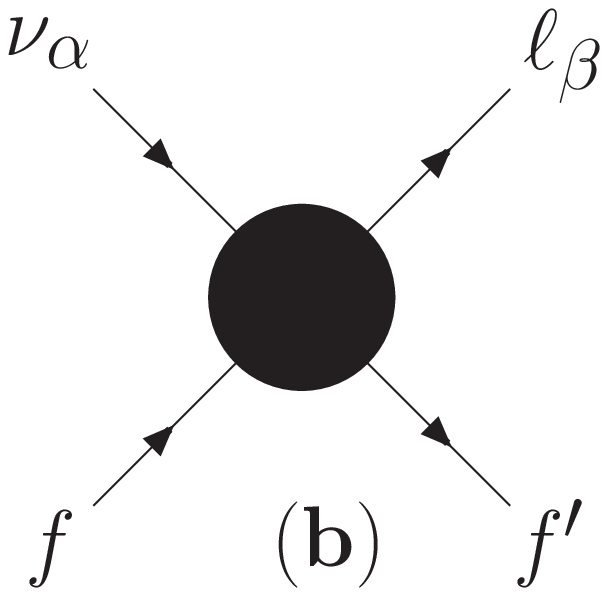,width=2.5cm}
\caption{Two types of the effective new interactions
which are relevant to neutrino oscillations.
}
\label{fig:newint}
\end{center}
\end{figure}

A class of effective non-standard neutrino interactions with
matter that would modify the neutrino oscillation probability
are given by
\begin{eqnarray}
{\cal L}_{\mbox{\rm\scriptsize eff}}^{\mbox{\tiny{\rm NSI}}} 
=\left\{\begin{array}{c}
-2\sqrt{2}\, \epsilon_{\alpha\beta}^{fP} G_F
(\overline{\nu}_\alpha \gamma_\mu P_L \nu_\beta)\,
(\overline{f} \gamma^\mu P f')~~(a)\\
-2\sqrt{2}\, {\epsilon'}_{\alpha\beta}^{fP} G_F
(\overline{\nu}_\alpha \gamma_\mu P_L \ell_\beta)\,
(\overline{f} \gamma^\mu P f')~~(b)
\end{array}\right.,
\label{NSIop}
\end{eqnarray}
which are depicted in
Fig.\,\ref{fig:newint}.  In Eq.\,(\ref{NSIop})
$f$ and $f'$ stand for fermions (the only relevant
ones here are electrons, u and d quarks),
$G_F$ is the Fermi coupling constant, $P$ stands for
a projection operator and is either
$P_L\equiv (1-\gamma_5)/2$ or $P_R\equiv (1+\gamma_5)/2$.
Since we are interested in the modification of
the neutrino oscillation phenomena due to new physics
here, the only relevant effective interactions
are four Fermi interactions of type Fig.\,\ref{fig:newint} (a) and (b),
which are neutral
and charged current interactions, respectively.
The presence of the interaction of
Fig.\,\ref{fig:newint} (a) would modify the matter effect during
propagation of neutrinos, while that of Fig.\,\ref{fig:newint} (b)
would change the process of production and detection of
neutrinos.

It has been shown \cite{Grossman:1995wx} by taking into account
various experimental constraints that the
absolute value of the coefficient ${\epsilon'}_{\alpha\beta}^{fP}$
of the interaction of type Fig.\,\ref{fig:newint} (b)
is small:
$|{\epsilon'}_{\alpha\beta}^{fP}|\lesssim{\cal O}(10^{-2})$.
On the other hand, in the case of Fig.\,\ref{fig:newint} (a),
it is known \cite{Davidson:2003ha,Friedland:2005vy} that the constraints
on ${\epsilon}_{\alpha\beta}^{fP}$ is relatively weak:
$|{\epsilon}_{\alpha\beta}^{fP}|\lesssim{\cal O}(1)$
for the flavor indices $\alpha,\beta=e,\tau$.
So in this talk I will consider
only new physics of type Fig.\,\ref{fig:newint} (a)
as a first step toward investigating new physics effects
at long baseline experiments.

\begin{figure}[h]
\begin{center}
\epsfig{file=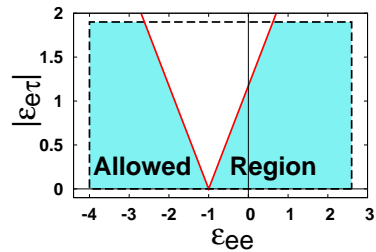,width=5cm}
\vglue -0.2cm
\caption{The allowed region (the shaded area)
in the $(\epsilon_{ee}, |\epsilon_{e\tau}|)$
plane.  Bounded by the dashed line is the region
allowed by various experimental data
\cite{Davidson:2003ha} and below the solid thick line is the
region allowed by the atmospheric neutrino and K2K
data \cite{Friedland:2005vy}.
}
\label{fig:np}
\end{center}
\end{figure}

In the presence of the new interaction of Eq.\,(\ref{NSIop}) (a),
by introducing the notation
$\epsilon_{\alpha\beta}
\equiv \sum_{P}
\left(
\epsilon_{\alpha\beta}^{eP}
+ 3 \epsilon_{\alpha\beta}^{uP}
+ 3 \epsilon_{\alpha\beta}^{dP}
\right)$, and by making the approximation
that the number density of electrons ($N_e$),
protons and neutrons are equal,
the $3\times3$ matrix of the matter potential becomes
\begin{eqnarray}
A\left(
\begin{array}{ccc}
1+ \epsilon_{ee} & \epsilon_{e\mu} & \epsilon_{e\tau}\\
\epsilon_{e\mu}^\ast & \epsilon_{\mu\mu} & \epsilon_{\mu\tau}\\
\epsilon_{e\tau}^\ast & \epsilon_{\mu\tau}^\ast & \epsilon_{\tau\tau}
\end{array}
\right),
\nonumber
\end{eqnarray}
where $A\equiv\sqrt{2}G_FN_e$.
From the analysis in \cite{Davidson:2003ha}
the coefficients involving the $\mu$ flavor are small:
$|\epsilon_{e\mu}| < 3.8\times 10^{-4}$,
$-0.05 < \epsilon_{\mu\mu} < 0.08$,
$|\epsilon_{\mu\tau}| < 0.25$.
So in the following discussions I will assume
$\epsilon_{e\mu}=\epsilon_{\mu\mu}=\epsilon_{\mu\tau}=0$
for simplicity
and keep in the analysis the remaining three parameters,
which have the values \cite{Davidson:2003ha}
$-4 < \epsilon_{ee} < 2.6$, $|\epsilon_{e\tau}| < 1.9$,
$|\epsilon_{\tau\tau}| < 1.9$.
Furthermore, it was shown in \cite{Friedland:2005vy}
that the atmospheric neutrino and K2K data imply
\begin{eqnarray}
|\epsilon_{e\tau}|^2
\simeq \epsilon_{\tau\tau} \left( 1 + \epsilon_{ee} \right),
\label{atm}
\end{eqnarray}
and $|\epsilon_{e\tau}|\lesssim |1 + \epsilon_{ee}|$.
Throughout the present talk I will assume that Eq.\,(\ref{atm})
holds exactly and eliminate $\epsilon_{\tau\tau}$ by Eq.\,(\ref{atm}).
Then we are left with
the two unknown parameters $\epsilon_{ee}$ and $\epsilon_{e\tau}$,
in addition to those in the standard framework.
Taking the constraints by \cite{Davidson:2003ha} and
\cite{Friedland:2005vy} into account,
the allowed region in the $(\epsilon_{ee}, |\epsilon_{e\tau}|)$
plane looks like Fig.\,\ref{fig:np}.
Below I will adopt the following
reference values for the oscillation parameters
in the standard three flavor framework:
$\Delta m^2_{31}=2.7\times10^{-3}$eV$^2$,
$\Delta m^2_{21}=8\times10^{-3}$eV$^2$,
$\sin^22\theta_{23}=1.0$,
$\sin^22\theta_{12}=0.8$.

\section{Analytic formula for the oscillation probability
$P(\nu_\mu\rightarrow\nu_e)$}

Before going into numerical analysis, it is instructive to
have an analytical expression of the oscillation probability
to see its behavior.
It was shown in \cite{Yasuda:2007jp} by generalizing
the exact analytical treatment on the oscillation probability
by Kimura--Takamura--Yokomakura \cite{Kimura:2002hb}
that the oscillation probability
$P(\nu_\mu\rightarrow\nu_e)$ in the presence of
the new interaction of Eq.\,(\ref{NSIop}) (a)
is obtained in the limit $\Delta m^2_{21}\rightarrow0$
as follows:
\begin{eqnarray}
\hspace{-5mm}
P(\nu_\mu\rightarrow\nu_e)&=&
-4\mbox{\rm Re}(\tilde{X}^{\mu e}_1\tilde{X}^{\mu e\ast}_3)
\sin^2\left[\frac{(\Lambda_+-\Lambda_-)L}{2}\right]
\nonumber\\
&-&
4\mbox{\rm Re}(\tilde{X}^{\mu e}_1\tilde{X}^{\mu e\ast}_2)
\sin^2\left(\frac{\Lambda_-L}{2}\right)
-4\mbox{\rm Re}(\tilde{X}^{\mu e}_2\tilde{X}^{\mu e\ast}_3)
\sin^2\left(\frac{\Lambda_+L}{2}\right)
\nonumber\\
&-&8\mbox{\rm Im}(\tilde{X}^{\mu e}_1\tilde{X}^{\mu e\ast}_2)
\sin\left(\frac{\Lambda_-L}{2}\right)\sin\left(\frac{\Lambda_+L}{2}\right)
\sin\left[\frac{(\Lambda_+-\Lambda_-)L}{2}\right],~~
\label{probnp}
\end{eqnarray}
where the energy eigenvalues $\Lambda_\pm$ and the
quantities $\tilde{X}^{\mu e}_j$ are given by
\begin{eqnarray}
\Lambda_\pm&=&(1/2)\left[\Delta E_{31}
+A(1+\epsilon_{ee})/\cos^2\beta
\right]\nonumber\\
&\pm&(1/2)
\sqrt{\left[\Delta E_{31}\cos2\theta_{13}''
-A(1+\epsilon_{ee})/\cos^2\beta
\right]^2
+(\Delta E_{31}\sin2\theta_{13}'')^2},
\nonumber\\
\tilde{X}^{\mu e}_1&=&
-[\xi+\eta e^{-i(\mbox{\rm\scriptsize arg}\,
(\epsilon_{e\mu})+\delta)}-\Lambda_+\zeta]
/[\Lambda_-(\Lambda_+-\Lambda_-)],
\nonumber\\
\tilde{X}^{\mu e}_2&=&
[\xi+\eta e^{-i(\mbox{\rm\scriptsize arg}\,
(\epsilon_{e\mu})+\delta)}-(\Lambda_++\Lambda_-)
\zeta]/(\Lambda_+\Lambda_-),
\nonumber\\
\tilde{X}^{\mu e}_3&=&
[\xi+\eta e^{-i(\mbox{\rm\scriptsize arg}\,(\epsilon_{e\mu})+\delta)}-\Lambda_-\zeta]
/[\Lambda_+(\Lambda_+-\Lambda_-)].
\nonumber
\end{eqnarray}
Here $\xi$, $\eta$, $\zeta$, $\beta$, $\theta''_{13}$, $\Delta E_{31}$
are given by
$\xi\equiv[(\Delta E_{31})^2+A(1+\epsilon_{ee})
\Delta E_{31}]U_{\mu3}|U_{e3}|$,
$\eta\equiv A\Delta E_{31}|\epsilon_{e\tau}|U_{\mu3}U_{\tau3}$,
$\zeta\equiv\Delta E_{31}U_{\mu3}|U_{e3}|$,
$\tan\beta\equiv|\epsilon_{e\tau}|/(1+\epsilon_{ee})$,
$\theta''_{13}=
\sin^{-1}| e^{-i\,\mbox{\rm\scriptsize arg}\,(\epsilon_{e\mu})}U_{e3}\cos\beta
+U_{\tau3}\sin\beta|$, $\Delta E_{31}\equiv\Delta m^2_{31}/2E$.

\begin{figure}[h]
\begin{center}
\epsfig{file=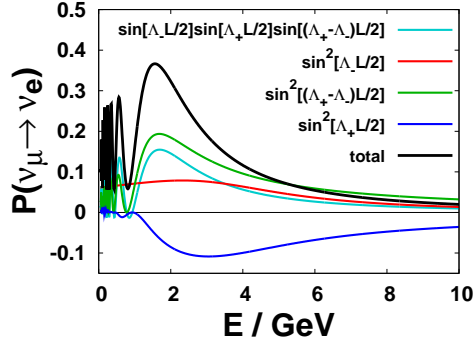,width=6.5cm}
\caption{Contribution of each term in Eq.\,(\ref{probnp})
to the oscillation probability
$P(\nu_\mu\rightarrow\nu_e)$ in matter at baseline $L$=730 km
in the presence of
new physics in the limit $\Delta m^2_{21}\rightarrow0$.
$\sin^22\theta_{13}=0.16$,
$\epsilon_{ee}=2.0$, $|\epsilon_{e\tau}|=1.5$,
$\mbox{\rm arg}(\epsilon_{e\tau})+\delta=\pi/2$
are assumed.}
\label{fig:prob2}
\end{center}
\end{figure}

Two remarks are in order.  First, Eq.\,(\ref{probnp}) indicates
that the phases appear in the probability only through
the combination of  $\mbox{\rm arg}\,(\epsilon_{e\mu})+\delta$ in the
limit $\Delta m^2_{21}\rightarrow0$.  It was found numerically
in \cite{Kitazawa:2006iq}
that this property holds approximately
even for nonvanishing $\Delta m^2_{21}$.  Secondly,
as is shown in Fig.\,\ref{fig:prob2}, each term in
Eq.\,(\ref{probnp}) gives a relatively large contribution,
and it is not easy to interpret the behavior of the probability
unlike in the standard three flavor case, where the probability
in the limit $\Delta m^2_{21}\rightarrow0$ can be expressed
by replacing the mixing angle $\theta_{13}$ and
the difference of the energy eigenvalues $\Delta E_{31}$ in vacuum
by those in matter, respectively \cite{Yasuda:1998sf}.

\section{Numerical analysis}

\begin{figure}[h]
\begin{center}
\epsfig{file=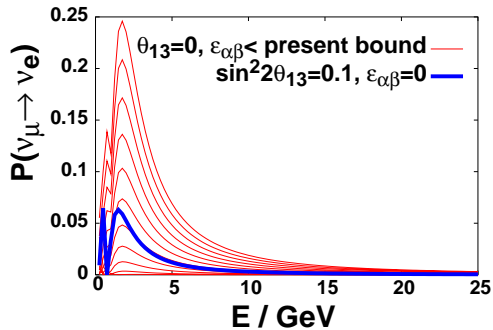,width=6.5cm}
\caption{The oscillation probability $P(\nu_\mu\rightarrow\nu_e)$
at the baseline L=730 km with (the thin solid lines)
or without (the thick solid line)
the new interaction of Eq.\,(\ref{NSIop}) (a) for various values of
$(\epsilon_{ee}, |\epsilon_{e\tau}|)$
within the allowed region in Fig.\,\ref{fig:np}.
}
\label{fig:prob1}
\end{center}
\end{figure}

In Fig.\,\ref{fig:prob1} the value of $P(\nu_\mu\rightarrow\nu_e)$
for the baseline $L$=730 km is plotted for various values of
$(\epsilon_{ee}, |\epsilon_{e\tau}|)$ in the allowed region depicted in
Fig.\,\ref{fig:np}, together with the value of the standard
case with nearly the maximum possible value
$\sin^22\theta_{13}=0.1$.
For some values of $(\epsilon_{ee}, |\epsilon_{e\tau}|)$
the oscillation probability becomes so large that it
cannot be explained by the standard three flavor framework.
We have done numerical analysis for two cases at the
MINOS experiments.

\begin{figure}[ht]
\begin{center}
\epsfig{file=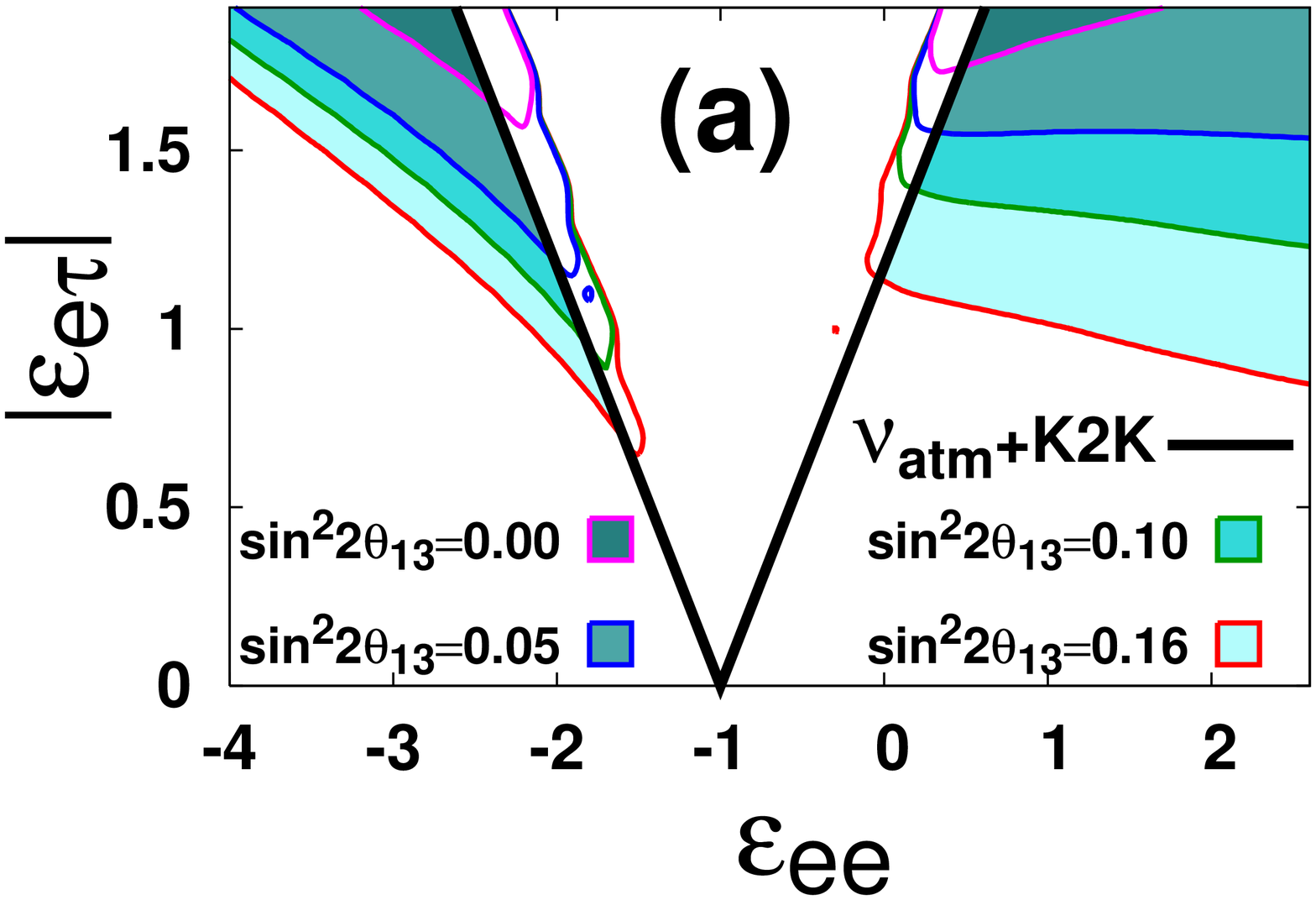,width=7.0cm}
\epsfig{file=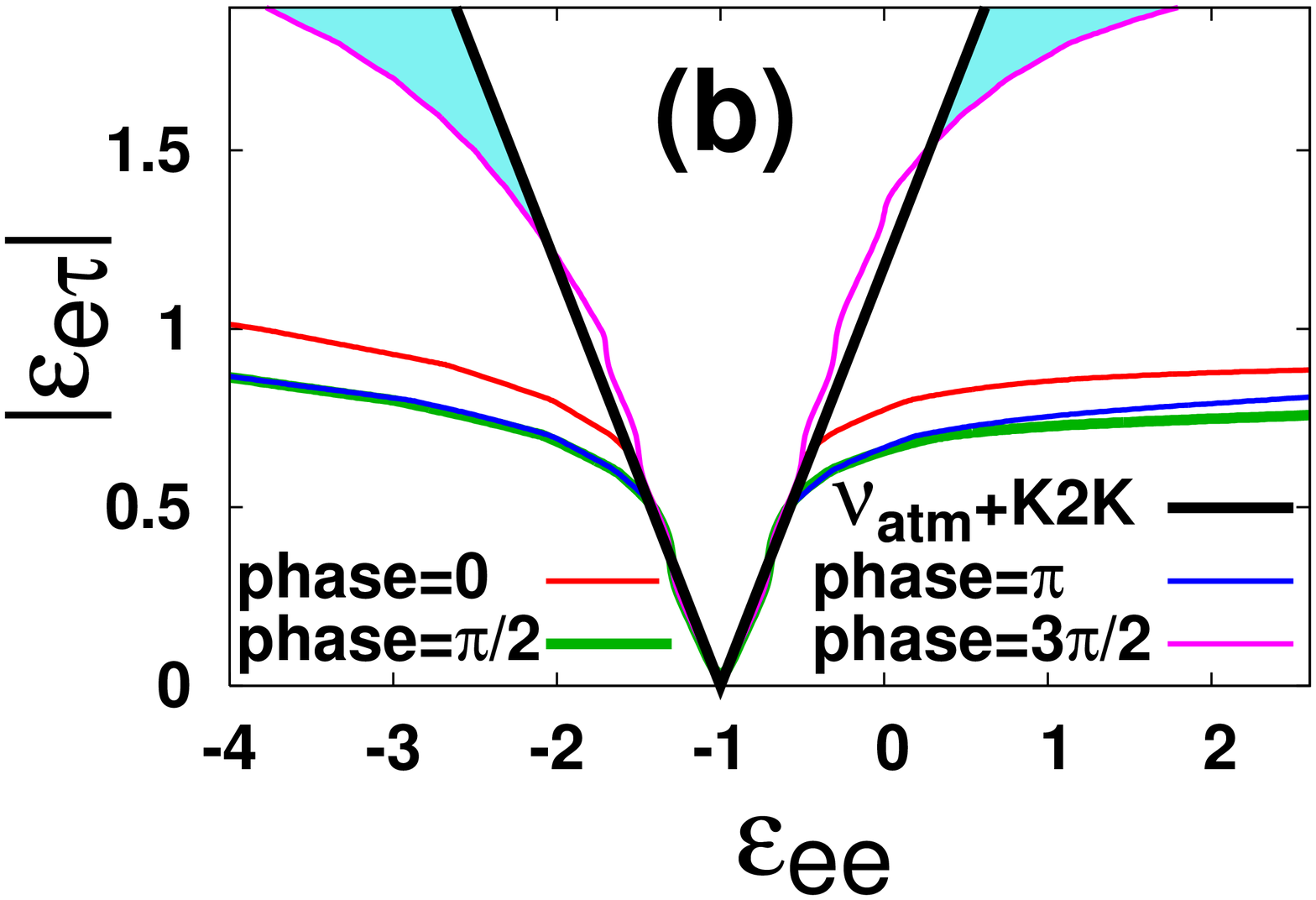,width=7.0cm}
\caption{Numerical results on sensitivity to new physics
by the appearance channel $\nu_\mu\rightarrow\nu_e$
at MINOS with $16\times10^{20}$ POT ($\simeq$ 5 years of running).  
(a) The shaded areas are the regions in the
$(\epsilon_{ee}, |\epsilon_{e\tau}|)$ plane
in which we can establish the existence of new physics
from the affirmative result of MINOS. 
The region varies, depending on the value of $\theta_{13}$.
(b) The shaded area is the region which can be excluded
from the negative result of MINOS.
The region depends on the value of the
phase $\mbox{\rm arg}(\epsilon_{e\tau})+\delta$.
Below the thick dashed line is
the region allowed by the atmospheric neutrino
and K2K data in both figures.
}
\label{fig:t}
\end{center}
\end{figure}

One is the case where
MINOS has an affirmative result for the
appearance channel $\nu_\mu\rightarrow\nu_e$.
There exists a certain region
in the $(\epsilon_{ee}, |\epsilon_{e\tau}|)$ plane
in which the difference between the numbers of events
with and without the new physics interaction
(the latter being the standard case with the maximum value
of $\sin^22\theta_{13}$) is so significant
that we can establish the existence of new physics.
The region in the $(\epsilon_{ee}, |\epsilon_{e\tau}|)$
plane for such a case depends on the value
of $\sin^22\theta_{13}$, and
is given by Fig.\,\ref{fig:t} (a).
The reason that a larger value of $\sin^22\theta_{13}$
gives a larger region is because the oscillation
probability is roughly additive in $\theta_{13}$
and $\epsilon_{\alpha\beta}$ so the larger value
$\theta_{13}$ has, the larger the number of events,
leading to the smaller statistical error and
the larger deviation from the standard case.
From Fig.\,\ref{fig:t} (a) we see that MINOS
potentially has a chance to establish
the existence of new physics, although
the region in the $(\epsilon_{ee}, |\epsilon_{e\tau}|)$
plane is relatively small for smaller values
of $\sin^22\theta_{13}$.

Another is the case where
MINOS has a negative result for $\nu_\mu\rightarrow\nu_e$.
In this case we can exclude a certain region
in the $(\epsilon_{ee}, |\epsilon_{e\tau}|)$ plane
whose prediction for the number of events is so large
that we have contradiction with the negative
assumption, irrespective of the value of $\theta_{13}$.
Such a region depends
on the value of $\mbox{\rm arg}(\epsilon_{e\tau})+\delta$,
and the case with
$\mbox{\rm arg}(\epsilon_{e\tau})+\delta=3\pi/2$
is the most pessimistic, i.e., the excluded
region becomes the smallest in this case.
As we can see from Fig.\,\ref{fig:t} (b),
again there is a little region in
the $(\epsilon_{ee}, |\epsilon_{e\tau}|)$ plane
which can be excluded by the negative result of MINOS.

\section{Conclusions}
As a first step in probing new physics
at long baseline experiments,
I have discussed
the new physics interaction given by Eq.\,(\ref{NSIop}) (a)
and have examined the sensitivity to such an interaction
by looking at the appearance channel $\nu_\mu\rightarrow\nu_e$
at the MINOS experiment.
In the process of the analysis, I presented the analytical
formula for $P(\nu_\mu\rightarrow\nu_e)$ which is exact
in the limit $\Delta m^2_{21}\rightarrow0$.
As far as the interaction Eq.\,(\ref{NSIop}) (a) is concerned,
an experiment with a longer baseline is more
advantageous since the new effect appears only through
the matter effect and roughly speaking it comes in
the oscillation probability in the form of
$\epsilon_{\alpha\beta}AL\sim\epsilon_{\alpha\beta}
(L/\mbox{\rm 2000 km})~(\alpha,\beta=e,\tau)$.
A neutrino factory in the future
\cite{ISS} will have much more statistics and
its baseline $L\gtrsim$ 3000 km gives larger sensitivity
to the matter effect,
so it is expected that a neutrino factory has
much better sensitivity to the new physics effect
discussed here.

\section{Acknowledgement}
The author would like to thank Janusz Gluza and
other organizers for invitation and hospitality
during the conference.
He would like to thank Hiroaki Sugiyama and
Noriaki Kitazawa for collaboration on \cite{Kitazawa:2006iq}.
He would also like to thank Chris Smith and the MINOS
collaboration for providing us the information of
their Monte Carlo study on the appearance channel.
This work was supported in part by a
Grant-in-Aid for Scientific Research of the Ministry of Education,
Science and Culture, \#19340062.

\end{document}